\documentclass[%
 aip,
 amsmath,amssymb,
 reprint,%
]{revtex4-1}

\usepackage{graphicx}
\usepackage{dcolumn}
\usepackage{bm}

\usepackage[utf8]{inputenc}
\usepackage[T1]{fontenc}
\usepackage{etoolbox}
\usepackage{amsmath}

\makeatletter
\def\@email#1#2{%
 \endgroup
 \patchcmd{\titleblock@produce}
  {\frontmatter@RRAPformat}
  {\frontmatter@RRAPformat{\produce@RRAP{*#1\href{mailto:#2}{#2}}}\frontmatter@RRAPformat}
  {}{}
}%
\makeatother
\begin{document}

\preprint{AIP/123-QED}

\title[]{Electrohydrodynamic flows inside a neutrally buoyant leaky dielectric drop}

\author{Joel R. Karp}
 \affiliation{INSA Rouen Normandie, Univ Rouen Normandie, CNRS, Normandie Univ, CORIA UMR 6614, F-76000 Rouen, France}

\author{Bertrand Lecordier}
\affiliation{INSA Rouen Normandie, Univ Rouen Normandie, CNRS, Normandie Univ, CORIA UMR 6614, F-76000 Rouen, France}

\author{Mostafa S. Shadloo}%
 \email{msshadloo@coria.fr}
\affiliation{INSA Rouen Normandie, Univ Rouen Normandie, CNRS, Normandie Univ, CORIA UMR 6614, F-76000 Rouen, France} 
\affiliation{Institut Universitaire de France, Rue Descartes, F-75231 Paris, France} 

\date{\today}

\begin{abstract}

We present for the first time an experimental investigation of electrohydrodynamic (EHD) flows within a neutrally buoyant drop with initial radius of 2.25 mm. Utilizing particle image velocimetry (PIV) and high-speed shadowgraphy, we measure the internal circulation and reported velocity profiles in the bulk and at the interface of the drop. Two leaky dielectric liquids, Silicone and Castor oils, are employed as the drop and external phase, allowing for the analysis of two shape configurations: oblate and prolate. The strength of the applied uniform electric field, $E_o$, spans from 0.125 to 1.75 kV/cm, enabling the analysis covering the small-deformation limit ($Ca_E \ll 1$), where the leaky dielectric model (LDM) is applicable. Drops with larger deformations, for which no analytical velocity field is available, are also investigated. Our measurements show a good agreement with the LDM theory for the small-deformation cases. The flows begin at the interface as a result of jump in the electric stresses, leading then to four counter-rotating vortices inside the drop. At permanent regime, the analytical solutions adequately predicts the radial and tangential velocity components, namely, $\mathbf{v_r}$ and $\mathbf{v_{\theta}}$, both in the bulk and at the interface of the drop. However, a nuanced behavior is noticed for larger deformations, where the LDM theory underpredicts the internal circulation. Moreover, due to the increased deformation, a non-uniform azimuthal profile is observed for the velocity at the interface, $\mathbf{v_{\theta}^i}$. Transient measurements of $\mathbf{v_{\theta}^i}$ enlighten the dynamic response of the EHD flows of the drop. Following the currently available analytical solutions, the dynamic response is governed by the time-scale of its deformation, $\tau_{def} = \mu a / \gamma$. We propose a critical value of electric capillary number $Ca_E$ of roughly 0.1 below which the LDM adequately describes the velocity field in both quasi steady-state and transitory regimes.

\vspace{5mm}


\end{abstract}

\maketitle

\section{Introduction}
\label{Section.Introduction}

Electrohydrodynamic (EHD) flows, driven by electric stresses shearing at fluid interfaces \cite{Vlahovska}, have garnered significant attention of the scientific community over the years. Specifically, drops dispersed in a liquid medium under a constant electric field serve as a pivotal reference case in the field, owing to their relevance in the pharmaceutical industry \cite{Elele}, drug delivery \cite{Iftimi}, and  in the energy sector \cite{Mousavi}, just to mention a few.

Consider a spherical drop with initial radius $a_o$ dispersed in another liquid with the same density ($\hat{\rho} = \rho$) but different viscosities ($\hat{\mu}$ and $\mu$), electrical conductivities ($\hat{\sigma}$ and $\sigma$) and permittivities ($\hat{\epsilon}$ and $\epsilon$). The hat symbol denotes a quantity associated to the drop phase. When an electric field with the strength of $E_o$ is applied, the discontinuity of electric properties at the interface generates a jump in electric stresses, $i.e.$ $\mathbf{T_e} - \mathbf{\hat{T_e}}$ \cite{Vlahovska}. This jump is balanced by hydrodynamic and capillary stresses, as described by:

\begin{equation}
   \mathbf{n} \cdot [(\mathbf{T} - \mathbf{\hat{T}})+(\mathbf{T_{e}} - \mathbf{\hat{T}_{e}})] = \gamma \mathbf{n} (\nabla \cdot \mathbf{n})
   \label{Eq.StressBalance}
\end{equation}

Here, $\gamma$ is the interfacial tension, and $\mathbf{n}$ is the unit vector normal to the interface. $T_{ij} = -p \delta_{ij} +\mu (\partial_j u_i + \partial_i u_j)$ are the hydrodynamic stresses where $\delta_{ij}$ is the Kronecker delta function. The electric stresses are calculated by the Maxwell stress tensor $T_{e_{ij}} = \epsilon (E_iE_j-E_kE_k \delta_{ij}/2)$. Equation \ref{Eq.StressBalance} implies that due to the discontinuities at the interface, there may be fluid motion both inside and outside of the drop, along with the deformation of its shape \cite{Abbasi2020}.

Several theoretical approaches have been proposed to quantify the electrohydrodynamic flows and the extent of drop deformation. Taylor (1966) \cite{Taylor} conducted a comprehensive analytical work considering both the drop and the external phases as poorly conducting leaky dielectric liquids. This led to the well-known leaky dielectric model (LDM). This model assumes small currents in the system and an irrotational electric field. If the charge relaxation time scale $\tau_c=\hat{\epsilon}/\hat{\sigma}$ is much smaller than other time scales, $e.g.$ the electrohydrodynamic flow time scale $\tau_{f}=\hat{\mu}/(\hat{\epsilon}E_o)$, the charge convection term can be neglected. Note that in this study focus is given to the circulation inside the drop and therefore the time scales are based on the properties of the drop phase. The transport equation of the charge then reduces to \cite{Saville2003}:

\begin{equation}
  \nabla \cdot \mathbf{E} = 0
  \label{Eq.delE}
\end{equation}

\noindent assuming a permanent regime and constant fluid properties. The fluid motion in the bulk of the drop phase is governed by the Stokes equations \cite{Melcher}, namely:

\begin{equation}
  \nabla \cdot \mathbf{\hat{v}} = 0
  \label{Eq.delV}
\end{equation}

\noindent and

\begin{equation}
 \hat{\mu} \nabla^2 \mathbf{\hat{v}} = \nabla p
 \label{Eq.Stokes}
\end{equation}

\noindent where $\mathbf{\hat{v}}$ is the fluid velocity vector and $p$ is pressure. Note that due to absence of charge convection, the electric forces are zero in the bulk and therefore are not present in the momentum balance. The velocity field inside the drop in permanent regime, can be obtained by solving these equations in polar coordinates \cite{Abbasi2020}:

\begin{equation}
\mathbf{\hat{v}_r} = \left[ \left( \frac{r}{a_o} \right)^3 - \left( \frac{r}{a_o} \right)\right] V_m \left( 1-3\cos^2 \theta \right)
\label{Eq.AnalyticalVR}
\end{equation}

\begin{equation}
\mathbf{\hat{v}_\theta} = \frac{1}{2} \left[ 5\left( \frac{r}{a_o} \right)^3 - 3\left( \frac{r}{a_o} \right)\right] V_m \sin2\theta
\label{Eq.AnalyticalVTHETA}
\end{equation}

\noindent where 

\begin{equation}
    V_m = \dfrac{9 \epsilon E_o^2 R (S - R)}{10 \mu(1+M)(R+2)^2}
    \label{Eq.MaxVel}
\end{equation} 

\noindent is the maximum velocity based on the conductivity ratio $R = \hat{\sigma}/\sigma$, the permitivitty ratio $S = \hat{\epsilon}/\epsilon$, and the viscosity ratio $M = \hat{\mu}/\mu$. 

The deformation parameter of the drop is defined as $D=(L-W)/(L+W)$, where $L$ and $W$ are the drop axes parallel and perpendicular to $\mathbf{E}$, respectively. The relaxation of the drop towards is equilibrium spherical shape is governed by the capillary or deformation time scale $\tau_{def}=\mu a_o / \gamma$. The relevant dimensionless number here is therefore the electric capillary number:

\begin{equation}
    Ca_E = \dfrac{\tau_{def}}{\tau_f} = \dfrac{\epsilon E_o^2 a_o}{\gamma},
    \label{Eq.ElectricCapillaryNumber}
\end{equation}

\noindent which was employed by \citet{Saville2003} to derive an analytical equation for the deformation parameter based on the LDM approach:

\begin{equation}
D_T = \frac{9 Ca_E}{16}f_T
\label{Eq.DT}
\end{equation}

\noindent where  

\begin{equation}
f_T = \frac{1}{(2+R)^2 \left[ R^2 + 1 -2S +3(R-S) \frac{2+3 \gamma}{5(1+\gamma)} \right]}
\label{Eq.FT}
\end{equation}

\noindent is the shape deformation function. The configuration of the drop is described by $R$ and $S$; if $R>S$ the internal circulation is from the equation to the poles, while the fluid motion is from the poles to the equator if $R<S$.

Although additional models have been proposed to account for the transitory dynamics \cite{Das2} \ \cite{Esmaeeli2011} \ \cite{Esmaeeli2020}, larger deformations \cite{Bentenitis2005} charge convection \cite{Sengupta2017}, and drop rupture \cite{Brosseau} \ \cite{Dubash} \ \cite{Karyappa2014}, the LDM proposed by Taylor \& Melcher \cite{Taylor} \ \cite{Melcher} remains a reference case for assessing the electrohydrodynamic flows of leaky dielectric drops in the small-deformation limit ($Ca_E << 1$), corroborated by diverse studies. Torza et al. (1971) \cite{Torza} were the first to report an experimental verification of the LDM obtaining a good qualitative agreement, with some comments on the conditions to rupture being also reported. Further studies were performed by Vizika \& Saville \cite{Vizika} and Ha \& Yang \cite{Ha2000} confirming the qualitative applicability of the LDM for the small-deformation limit, but commenting on the necessity of improvements to account for larger deformations. In summary, it is understood that the pioneer LDM theory is capable of describing the EHD dynamics of drops fairly well, considering the small-deformation limit and permanent regime. For a more in-depth review of this topic, readers are directed elsewhere \cite{Saville2003} \ \cite{Vlahovska} \ \cite{Abbasi2020}. From an experimental standpoint, the majority of investigations have focused on the shape of the drop \cite{Emilij2002} \ \cite{Abbasi2019} \ \cite{Jiang2020}, on its coalescence \cite{Huang2019} and on its breakup \cite{Dong}. For the latter case, the conductivity and permittivity ratios dictate different modes of breakup. For a strong electric field and leaky dielectric fluids, if the drop is more conducting and viscous, the tangential stresses lead to the emission of jets. Otherwise, if the fluid system is inverted, a lenticular drop is formed leading to equatorial streaming \cite{Wagoner2021}. Moreover, a non-uniform velocity field can be observed for a certain range of electric and hydrodynamic properties of the fluids, leading to the rotation of the drop \cite{Salipante2010}. \citet{Mikkelsen2018} were the first to report experimental measurements of the electrohydynamic flows of a drop; using particle image velocimetry (PIV), they analyzed the suppression of the velocity field by Marangoni flows due to surface contaminants. However, a detailed analysis of the velocity field, particularly close to the interface, nor the internal circulation was reported.  

More recently, the focus has shifted towards the investigation of the transient dynamics of drops under a constant electric field \cite{Sherwood} \ \cite{Dubash} \ \cite{Mandal} \ \cite{Das}. A first theoretical analysis of the dynamic response of the EHD flows was presented by \citet{Sozou}. Assuming that the charge instantaneously accumulate at the interface, the velocity field was investigated with time within the creeping flow regime considering a low electric Reynolds number $Re_E = \rho a_o v_s / \mu$ where $v_s = \epsilon E_o^2 a_o / \mu$ is the characteristic velocity. Thus, once the steady state is reached the flow converged to the LDM solution; however, no closed-form solution of the transient velocity field was proposed. To fill this gap, Esmaeeli \& Sharifi (2011) \cite{Esmaeeli2011} followed a different approach by neglecting the local fluid acceleration term $\partial \mathbf{v} / \partial t$ also considering creeping flow. Despite an analytical solution of the velocity field being outlined, the authors focused on the deformation-time curve of the drop with little discussion on the dynamics response of the velocity field. Moreover, a monotonic behavior was proposed, with a modified the capillary time scale governing the dynamics. Later on, it was shown by \citet{Esmaeeli2020} that the monotonic behavior adequately describes the dynamics of a slightly-deformed drop with small Ohnesorge numbers, $i.e.$ $Oh = (\mu a_o / \gamma) / (a_o^2 \rho / \mu) < 1$, whereas an oscillatory time response is observed when the time scale of momentum diffusion is larger for $Oh>1$. \citet{Lanauze2015} extended the solution of \citet{Sozou} by considering charge relaxation under creeping flow using a boundary integral method. A similar approached has been performed by \citet{Das2} who developed a second-order perturbation solution to include both charge relaxation and charge convection in the creeping glow regime.

Based on the literature review outlined above, experimental inquiries into the internal circulation of a leaky dielectric droplet are to best of our knowledge not available. Even within the small-deformation limit at permanent regime, where the LDM theory applies fairly, the experimental validation of the LDM model is thus far confined to the deformation of the drop. As for the dynamic response of the drop, existing experimental studies have primarily focused on the shape of the drop, while a proper quantification of fluid circulation is, to the best of our knowledge, restricted to numerical and theoretical investigations. Moreover, the definition of a "small-deformation limit" remains vague to date. A criterion that quantitatively determines the extent of the applicability of the LDM approach is yet to be reported. 

Therefore, experimental measurements of the EHD circulation within a neutrally buoyant drop are still necessary for a full comprehension of the phenomenon, particularly regarding the temporal evolution of the velocity field outside the small-deformation limit for which no investigations are reported. In this study, we propose a novel methodology based on particle image velocimetry (PIV) to quantify the internal circulation of neutrally buoyant drop under a constant electric field. The methodology provides accurate measurements of the velocity field inside the drop for a wide range of $Ca_E$.

\section{Experimental methodology}
\label{Section.Methods}

\subsection{Set-up and test fluids}

\begin{figure*}
  \centerline{\includegraphics[width = \textwidth]{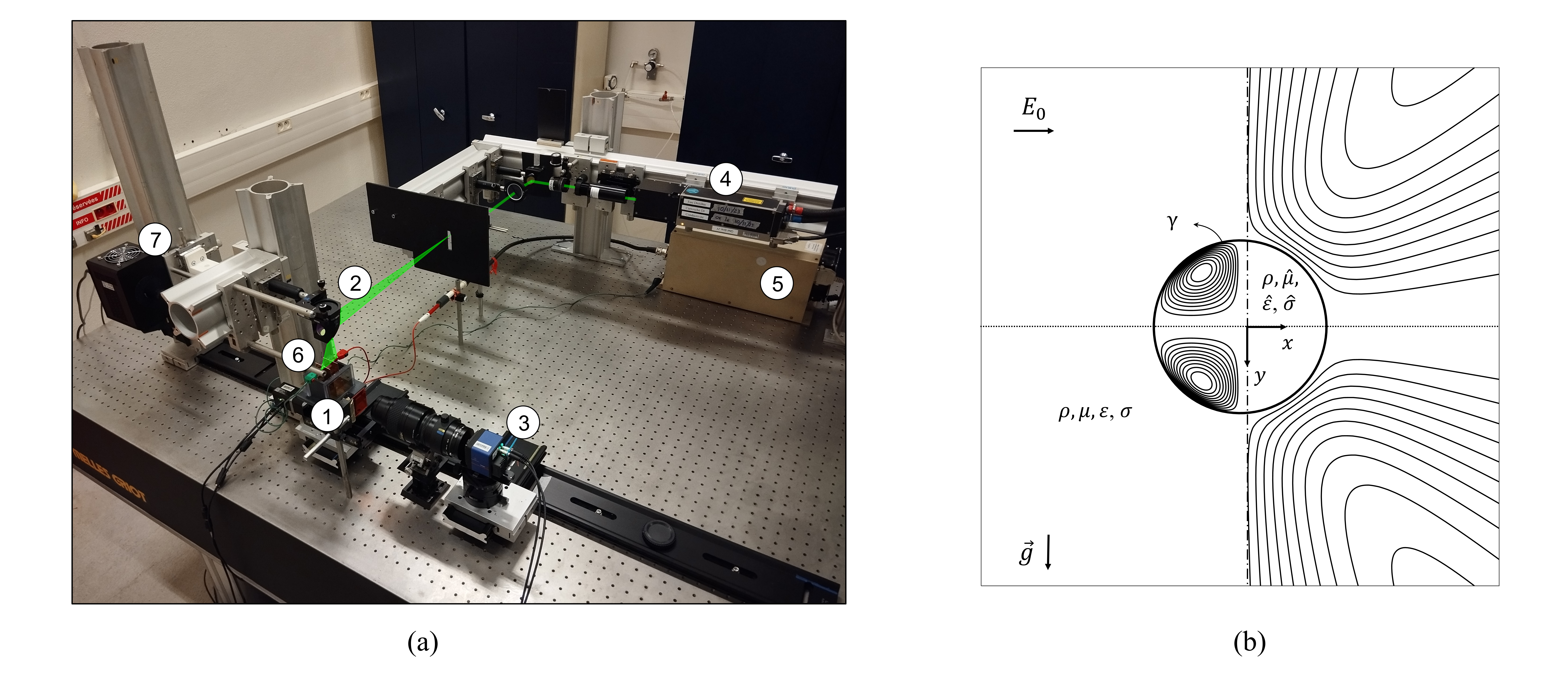}}
  \caption{(a) Experimental apparatus for measuring internal electrohydrodynamic flows in a drop showing the cuvette (1), laser sheet (2), CCD camera (3), laser source (4), high-voltage power supply (5), electrode system (6), and background light source (7). (b) illustration of the fluid system.}
\label{Fig.Setup}
\end{figure*}

The experiments are conducted in a plexiglass cuvette with a base of $50 \times 40$ mm$^2$ and $40$ mm of height (Fig. \ref{Fig.Setup}a). The cuvette is completely filled with a stagnant leaky dielectric liquid, where a drop of $a_o = 2.25$ mm is inserted using a micropipette. The size of the drop is kept fixed for all measurements. Silicone oil (8453.290, VWR) and Castor oil (24667.290, VWR) are either the continuous medium or the drop, depending on the case analyzed. Table \ref{Tab.Properties} summarizes the properties of the fluids. The electrical permittivities and conductivities are measured using broadband dielectric spectroscopy (Novocontrol-concept 80, Germany) following the parallel electrode method \cite{Araujo}. The density, dynamic viscosity, refractive indexes, and the interfacial tension are obtained from the supplier at 20 \textdegree C (the temperature control is ensured by using a ventilation system). The electric field is generated using two vertical copper electrodes with thickness of 1 mm spaced 20 mm apart. A high-voltage supply system (Hartlauer Elektronik CC400-32, Germany) is used to apply the desired voltage. Figure \ref{Fig.Setup} illustrates the experimental set-up, including a sample image of the internal and external circulation of the drop for visualization.

Table \ref{Tab.Cases} presents the cases examined in this study. By utilizing the same fluids and modifying the composition of both the continuous phase and the drop, we achieve two distinct final shapes: the oblate (Cases I to IV) and the prolate (Cases V to VIII) configurations. A drop is considered oblate or prolate when its major axis is perpendicular or parallel to the direction of the electric field, $\mathbf{E}$, respectively. The strength of the electric field is $E_o = \Delta V / \Delta x$ where $\Delta V$ is the voltage and $\Delta x$ is the distance between the electrodes. $E_o$ spans from 0.25 to 1.75 kV/cm and from 0.125 to 1.10 kV/cm for the oblate and prolate configurations, respectively. Therefore, $D$ is kept nearly constant when the fluid composition is inverted. The relevant dimensionless numbers are also presented in Tab. \ref{Tab.Cases}. 

\begin{table*}
\begin{center}
\def~{\hphantom{0}}
\includegraphics[scale = 0.38]{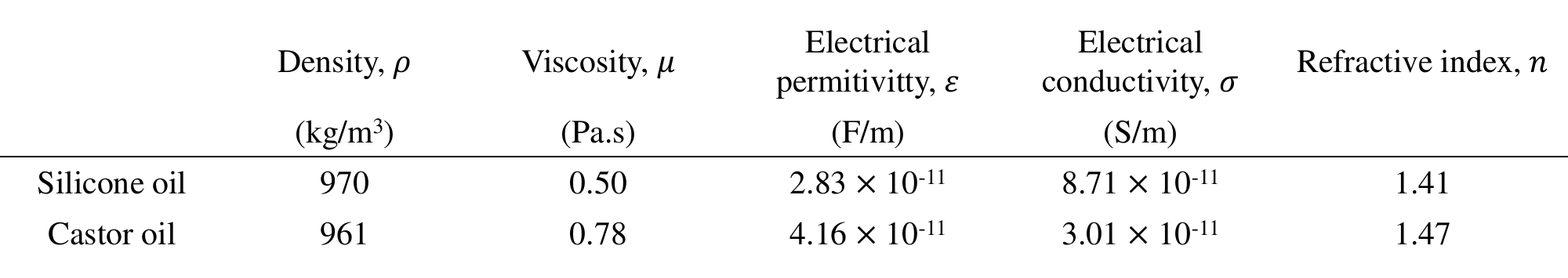}
\caption{Properties of the working fluids employed in this study The interfacial tension is $\gamma = 4$ mN/m.}
\label{Tab.Properties}
\end{center}
\end{table*}

\begin{table*}
\begin{center}
\def~{\hphantom{0}}
\includegraphics[scale = 0.38]{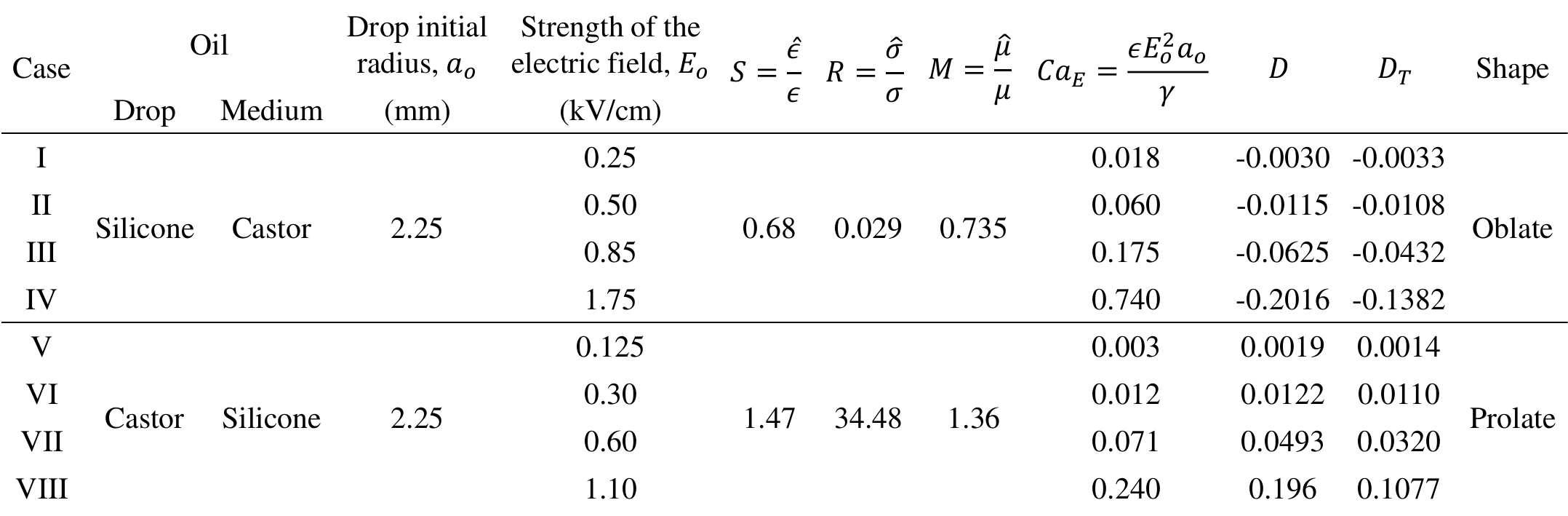}
\caption{Cases analyzed in this study, indicating the fluids serving as the drop and the medium, along with the initial drop radius ($a_0$ in millimeters) and electric field strength ($E_o$). The table also presents permitivity, conductivity, and viscosity ratios, the electric electric capillary number $Cae_E$ and the deformation parameters $D$ (obtained experimentally) and $D_T$ (Eq. \ref{Eq.DT}).}
\label{Tab.Cases}
\end{center}
\end{table*}

\subsection{High speed shadowgraphy}

\begin{figure*}
  \centerline{\includegraphics[width = \textwidth]{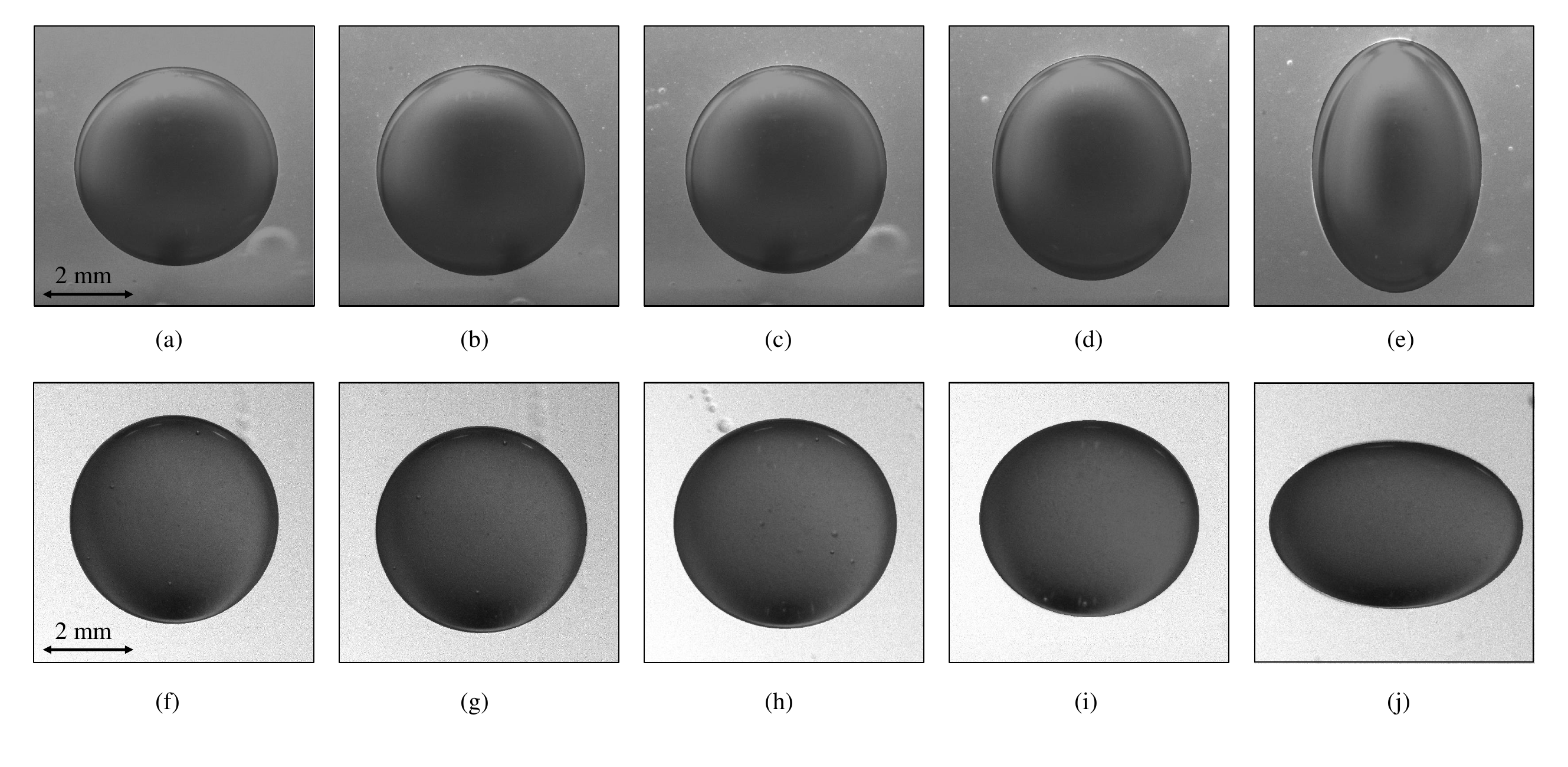}}
  \caption{Shadowgraph images for a silicone drop dispersed in Castor oil (top row) : no electric field (a), Cases I (b), II (c), III (d), and IV (e), and for a Castor oil drop dispersed in silicone oil (bottom row) : no electric field (f), Cases V (g), VI (h), VII (i), and VIII (j).}
\label{Fig.ShadowImages}
\end{figure*}

High speed shadowgraphy measurements are set up according to the experimental set-up shown in Fig. \ref{Fig.Setup}. The contour of the drop is quantified using a single CCD camera (CX3-25, LaVision, Germany) with a spatial resolution of 5296 x 4584 pixels with 12 bits of digital output, coupled to a 105 mm lens (Nikkon, Japan). A LED system is used as a background source illumination. The shadowgraph images of the drop are then recorded with a frequency of 20 Hz. Figure \ref{Fig.ShadowImages} shows sample images of the drop, at steady-state condition, for the cases shown in Tab. \ref{Tab.Cases}. Clearly, the drop evolves from a perfectly spherical shape when no electric field is applied (Fig. \ref{Fig.ShadowImages} (a) and \ref{Fig.ShadowImages} (f)) to either an oblate (top row) or a prolate (bottom row) shape, depending on the fluid composition of the system. Moreover, the extend of the shape deformation evolves with $E_o$. From the shadowgraph images shown in Fig. \ref{Fig.ShadowImages}, it is possible to calculate the deformation parameter $D$ as presented in Tab. \ref{Tab.Properties} for each case investigated. A robust agreement is observed between $D$ and $D_T$ calculated from Eq. \ref{Eq.DT}, for Cases I \& II and and V \& VI when $Ca_E \ll 1$. However, a large deviation is noticed when we consider drops with larger deformations (Cases III, IV, VI, VII, and VIII) where Eq. \ref{Eq.DT} is no longer valid to estimate $D$ properly. Finally, we stress that although $\hat{\rho}$ is slightly different than $\rho$, buoyancy effects are apparently negligible based on the good agreement with the analytical solution.

\subsection{Particle image velocimetry}

The internal circulation of the drop is quantified by adding RhodamineB tracer particles (20 $\mu$m) and illuminating the drop at its plane of symmetry with a laser sheet (Fig. \ref{Fig.Setup}). A monocavity Nd:YAG laser (Quantel BRIO SP23, France) with a wavelength of 532 nm and energy of 64 mJ served as the laser source. Slow recirculation within the drop allowed image acquisition at a relatively low frame rate (20 Hz), enabling the use of a monocavity laser instead of a double-cavity one. Mirrors, spherical lenses, and cylindrical lenses are arranged to generate a laser sheet approximately 150 $\mu$m thick. The image acquisition system for the PIV measurement is the same used from the shadowgraph experiments, as shown previously.  

\begin{figure*}
  \centerline{\includegraphics[width = \textwidth]{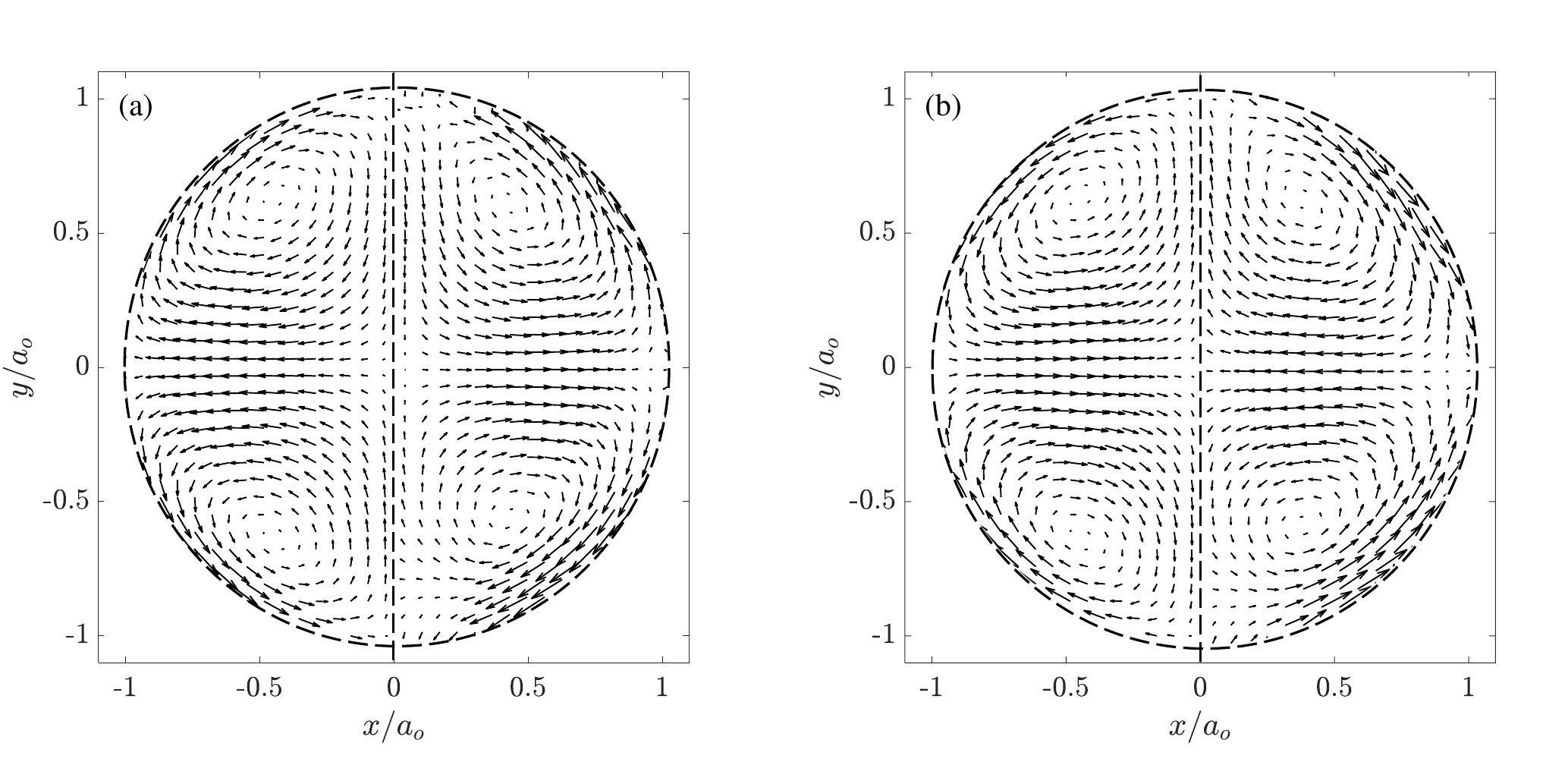}}
  \caption{Comparison between the analytical solution given by Eqs. \ref{Eq.AnalyticalVR} and \ref{Eq.AnalyticalVTHETA} (left half of each sub-figure) and the velocity vector obtained experimentally (right half of each sub-figure) for Cases II (a) and V (b).}
\label{Fig.VectorsPIV}
\end{figure*}

The vector fields shown in Fig. \ref{Fig.VectorsPIV} (right half of each sub-figure) for Cases II and V are obtained using the Davis software (version 10, LaVision, Germany) with a cross-correlation iterative multi-pass routine of round interrogation windows of 64 $\times$ 64 pixels with 50 \% overlap in both directions. The final velocity field is then obtained after a correlation-based validation to remove false vector outside the drop. The velocity vectors are then corrected based on the refraction of the light emitted by the particles as it crosses the drop interface. Details of this simplified procedure are shown in Appendix A. Figure \ref{Fig.VectorsPIV} provides a visual comparison between the PIV velocity fields and the analytical velocity vectors (left half of each sub-figure) given by Eqs. \ref{Eq.AnalyticalVR} and \ref{Eq.AnalyticalVTHETA}, showing a good qualitative agreement. 

\section{Results and discussion}

\begin{figure*}
  \centerline{\includegraphics[width = \textwidth]{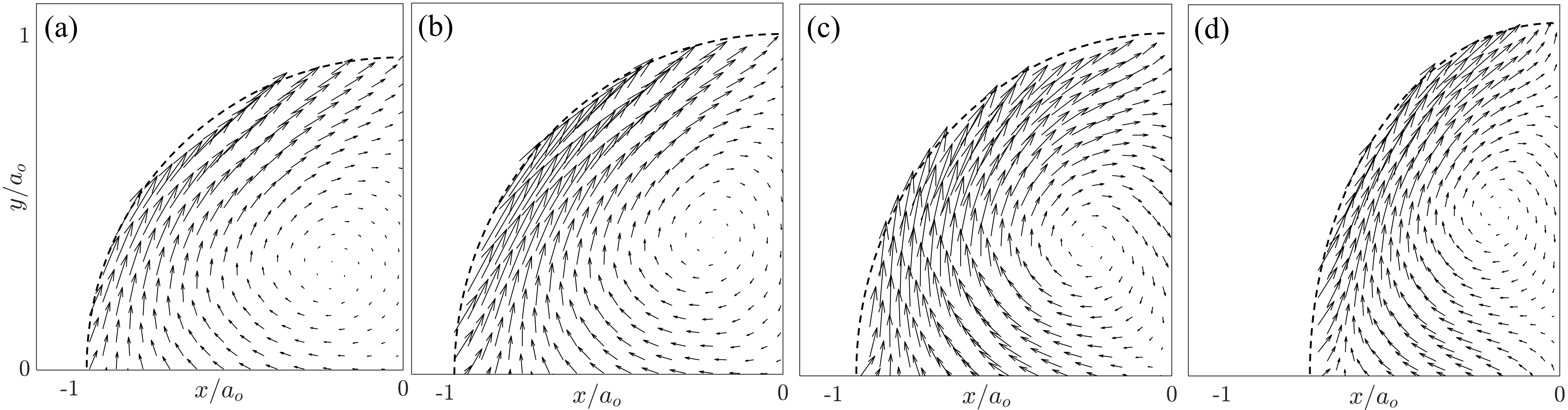}}
  \caption{Velocity vectors measured experimentally for Cases I (a), II (b), III (c), and IV (d). The dashed line shows a visual representation of the interface of the drop. Fewer vectors are shown for visualization purposes.}
\label{Fig.VectorsPIV14}
\end{figure*}

\begin{figure*}
  \centerline{\includegraphics[width = \textwidth]{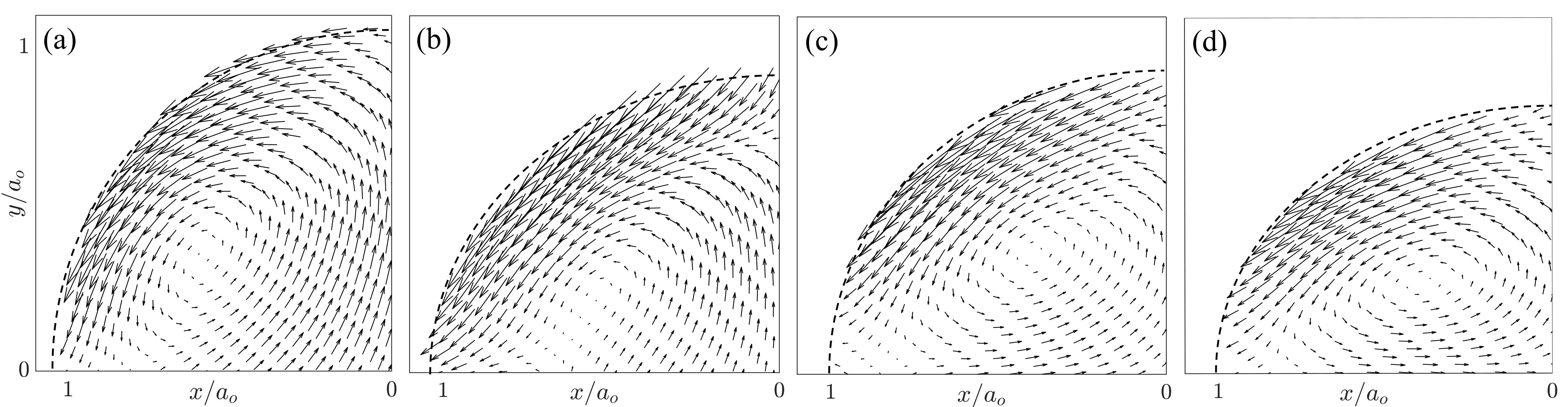}}
  \caption{Velocity vectors measured experimentally for Cases V (a), VI (b), VII (c), and VIII (d). The dashed line shows a visual representation of the interface of the drop. Fewer vectors are shown for visualization purposes.}
\label{Fig.VectorsPIV58}
\end{figure*}

Figures \ref{Fig.VectorsPIV14} and \ref{Fig.VectorsPIV58} show the measured velocity vectors for the cases shown in Tab.  \ref{Tab.Cases}, considering permanent regime. For the sake of clarity, the vectors are shown for the upper left quadrant only that corresponds to $-1<x/a_o<0$ and $0<y/a_o<1$. By making the positions dimensionless as a function of the initial radius $a_o$, a visual inspection of the shape of the drop can also be performed. Note  that the dashed line shown in the figures qualitatively represent the interface of the drop. An observation is possible from an inspection of Cases I and II: the velocity field remains symmetrical when the deformation of the drop is small. For these cases, the velocity field is seemingly symmetrical with respect to the diagonal, $i.e.$ $x/a_o = y/a_o$. The circulation is bounded at $x/a_o = y/a_o=0$, while the tangential component $\mathbf{v_{\theta}}$ is maximum at the diagonal crossing the quadrant, close to the interface. A stagnation zone is also identified at $x/a_o=-0.2$ \& $y/a_o=0.3$ for Case I and at $x/a_o=-0.2$ \& $y/a_o=0.4$ for Case II. The slight deviation regarding the horizontal prediction is consistent with the oblate shape, as shown previously in Tab. \ref{Tab.Cases}. However, the larger deformations observed for Cases III and IV lead to a non-uniform velocity field, thus shifting the center of the circulation zone to $x/a_o=-0.2$ \& and $y/a_o=0.5$ and $x/a_o=-0.2$ \& $y/a_o=0.6$, respectively. Clearly, the symmetry axis along the diagonal line of the quadrant is lost for these two cases. When the composition of the fluids is reversed, $i.e.$ Cases V to VIII, a similar observation is made regarding the evolution of the velocity vectors with $E_o$: a non-uniform distribution is noticed when $E_o$ is larger, particularly Cases VII and VIII. Under those circumstances, it is expected that the analytical velocity field (Eqs. \ref{Eq.AnalyticalVR} and \ref{Eq.AnalyticalVTHETA}) and shown in Fig. \ref{Fig.VectorsPIV} no longer matches our measurements.

\subsection{Radial velocity profiles}

\begin{figure*}
  \centerline{\includegraphics[width = \textwidth]{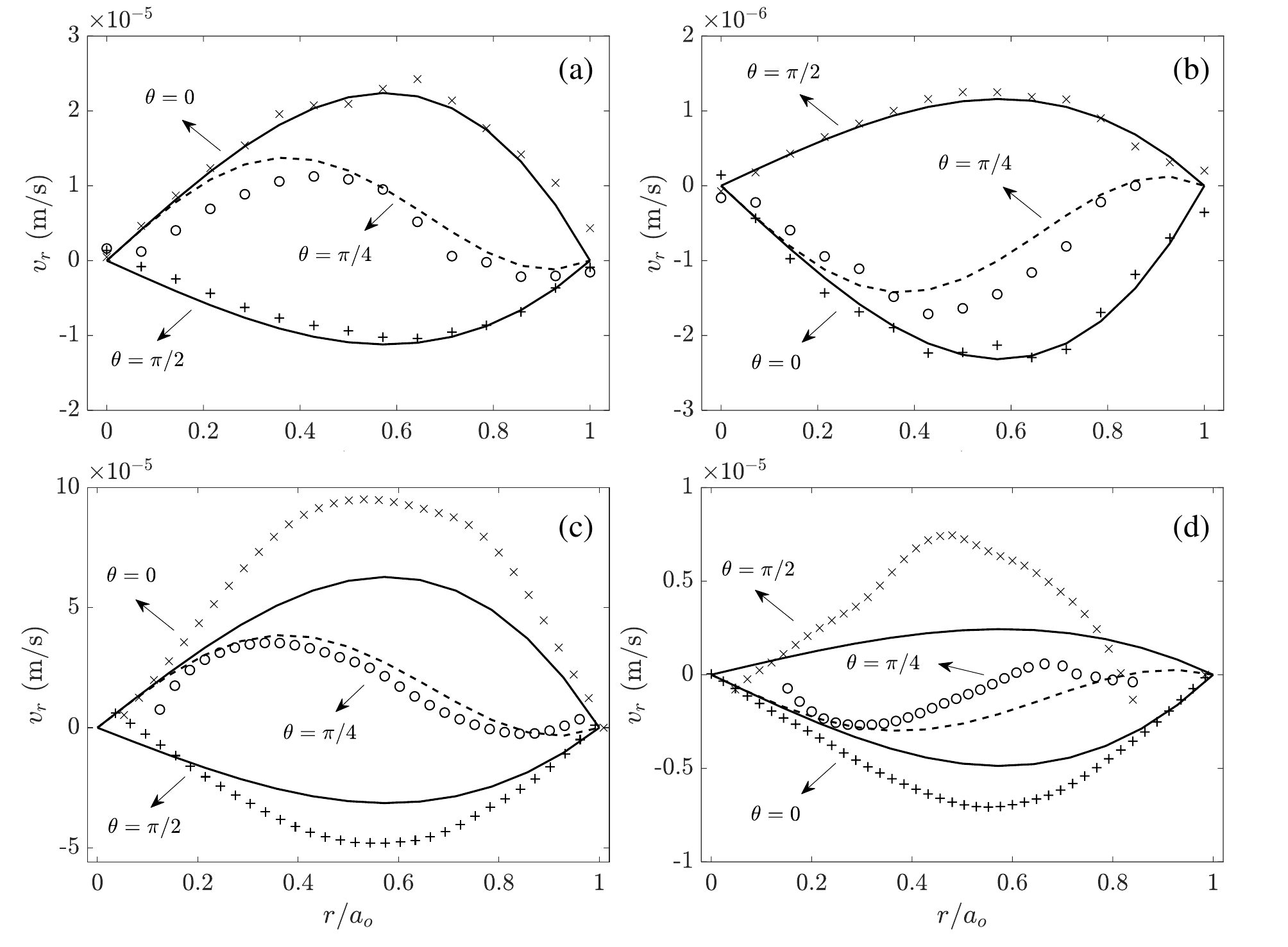}}
  \caption{Profiles for the radial velocity component $\mathbf{v_r}$ for $\theta = 0$, $\theta = \pi/4$ and $\theta=\pi/2$ for Cases II (a), V (b), III (c), and VI (d). Solid lines represent the radial profiles obtained from the analytical velocity field (Eq. \ref{Eq.AnalyticalVR}) and markers show the experimental results for the same value of $\theta$.}
\label{Fig.RadialVelocityProfile}
\end{figure*}

Figure \ref{Fig.RadialVelocityProfile} shows radial profiles of $\mathbf{v_r}$. The spatial coordinates are now converted from the Cartesian ($x,y$) to the polar ($r,\theta$) system to improve the clarity of the results. The radial profiles are shown along the $r$ distance for three different values of $\theta$, namely, $0$, $\pi/4$ and $\pi/2$. Let us consider Cases II and VI shown in Figs. \ref{Fig.RadialVelocityProfile} (a) and (b), respectively. Clearly, the figure suggests a very good agreement between the experimental measurements (shown in the figure as the markers) and the analytical profiles obtained (lines in Fig \ref{Fig.RadialVelocityProfile}). Considering the profiles at the axes of symmetry of the drop, $i.e.$ $\theta=0$ and $\theta=\pi/2$, a parabolic behavior (slightly inclined towards the region where $r/a_o \rightarrow 1$) is present for both cases analyzed. At $\theta=\pi/4$ $\mathbf{v_r}$ oscillates due to the presence of an additional stagnation point, located at the center of the vortex, roughly at $r = 0.7$. 

A different $\mathbf{v_r}$ profile is observed when we shift our analysis to more deformed drops, $e.g.$ Cases III and VII. According to Eq. \ref{Eq.MaxVel}, the contribution of $E_o$ is solely regarding the magnitude of the velocity components, as indicated in Fig. \ref{Fig.RadialVelocityProfile} (c) and (d) that show the exact same profile of $\mathbf{v_r}$ from the LDM theory. On the other hand, this clearly deviates from our measurements, as the profiles of $\mathbf{v_r}$ no longer match the analytical prediction. Although for Case III a similar profile is still observable, the theory underpredicts the magnitude of $\mathbf{v_r}$. The deviation is particularly significant for Case VII (shown in Fig \ref{Fig.RadialVelocityProfile} (d)), where the profile obtained experimentally is completely assymmetric. Strictly speaking, the LDM applies only $Ca_E<<1$, which was the case shown in Fig. \ref{Fig.RadialVelocityProfile} (a) and (b). A remark can be drawn here to the shape of the drop. \citet{Abbasi2020} reported that a Castor oil drop dispersed in Silicone oil may rupture when $Ca_E$ is roughly 0.2. However, our measurements suggest that before this limit deviations in the internal circulation are already evident when $Ca_E \approx 0.1$.

\subsection{Quasi-steady velocity at the interface}

\begin{figure*}
  \centerline{\includegraphics[width = \textwidth]{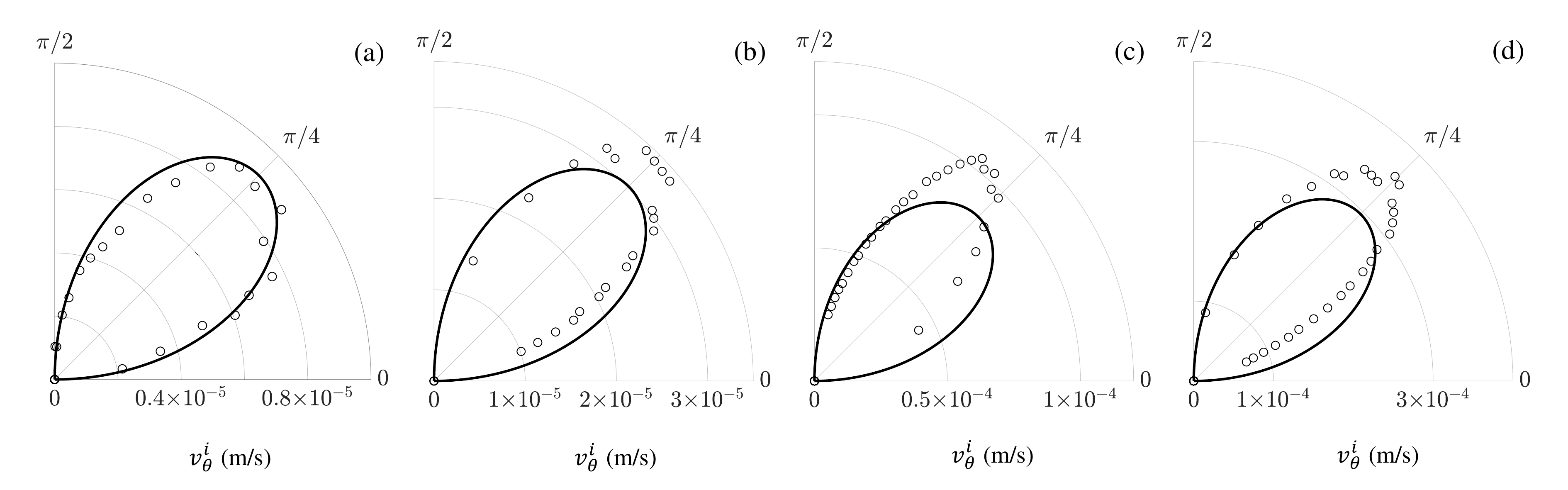}}
  \caption{Markers indicate the tangential component of the velocity at the drop interface for Cases I (a), II (b), III (c), and IV (d) in permanent regime. The solid line represents the analytical solution given by Eq. \ref{Eq.AnalyticalVTHETA}.}
\label{Fig.TangentialVelocityInterface14}
\end{figure*}

\begin{figure*}
  \centerline{\includegraphics[width = \textwidth]{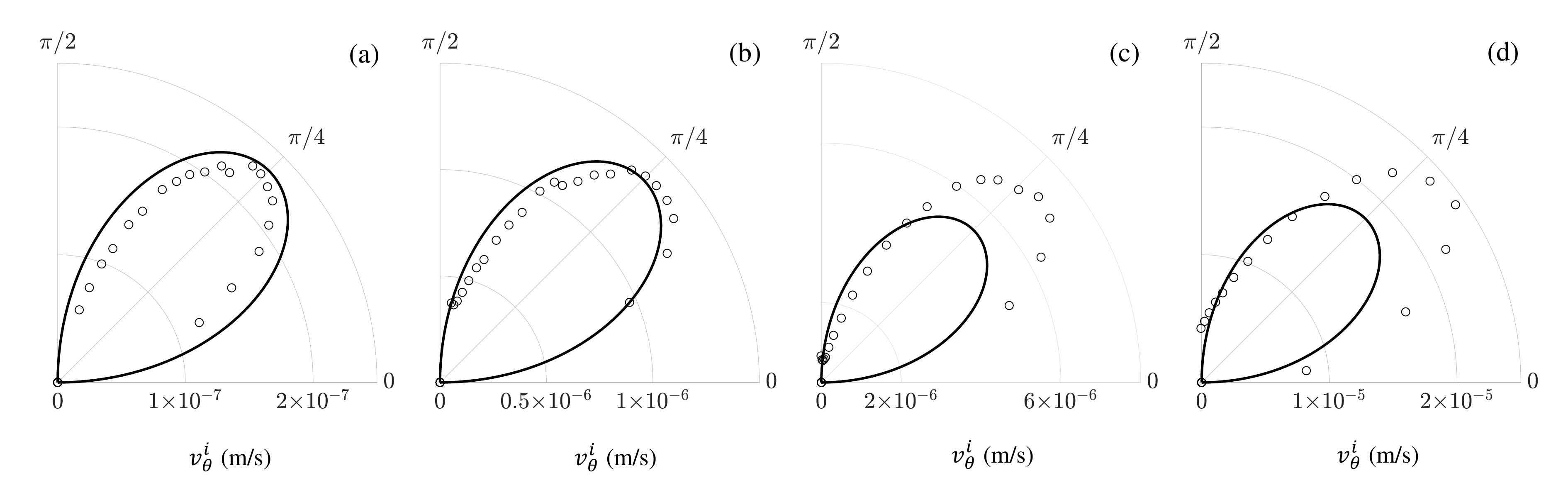}}
  \caption{Markers indicate the tangential component of the velocity at the drop interface for Cases V (a), VI (b), VII (c), and VIII (d) in permanent regime. The solid line represents the analytical solution given by Eq. \ref{Eq.AnalyticalVTHETA}.}
\label{Fig.TangentialVelocityInterface58}
\end{figure*}

The tangential component of the velocity, $\mathbf{v_{\theta_i}}$, at the drop interface is shown in Figs. \ref{Fig.TangentialVelocityInterface14} and \ref{Fig.TangentialVelocityInterface58} for the cases shown in Tab. \ref{Tab.Cases}. The measurements correspond to all the velocity fields shown in Figs. \ref{Fig.VectorsPIV14} and \ref{Fig.VectorsPIV58}. The experimental values (indicated as markers) closely align with the analytical solution predicted by Eq. \ref{Eq.AnalyticalVTHETA} for the cases shown shown in Figs. \ref{Fig.TangentialVelocityInterface14} and \ref{Fig.TangentialVelocityInterface58} (a) and (b). However, it is noteworthy that there is a slight non-uniform trend of the azimuthal profile of $\mathbf{v_{\theta_i}}$ even within the small-deformation limit. Mainly, the azimuthal profiles tends to be thinner and slightly inclined towards $\theta \rightarrow 0$; a probable consequence of the slight deformation observed even for those cases, which is not all taken into account by the LDM model. However, such deviation is small and a good agreement can be considered here. Again, the measurements of $\mathbf{v_{\theta_i}}$ evolve to a non-uniform azimuthal profile when $E_o$ is further increased, as shown in Fig. \ref{Fig.TangentialVelocityInterface14} and Figs. \ref{Fig.TangentialVelocityInterface58} (c) and (d). Besides the significant underprediction of our measurements by the LDM model, the polar coordinate $\theta$ when $\mathbf{v_{\theta_i}}$ is maximum, $i.e.$ $\mathbf{v_{\theta_i}}^m$ moves towards the upper ($\theta \rightarrow \pi/4$) and lower ($\theta \rightarrow 0$) regions of the quadrant, respectively, for Cases III to IV, and VII to VIII. 

\subsection{Dynamic response of the velocity at the interface}

\begin{figure*}
  \centerline{\includegraphics[width = \textwidth]{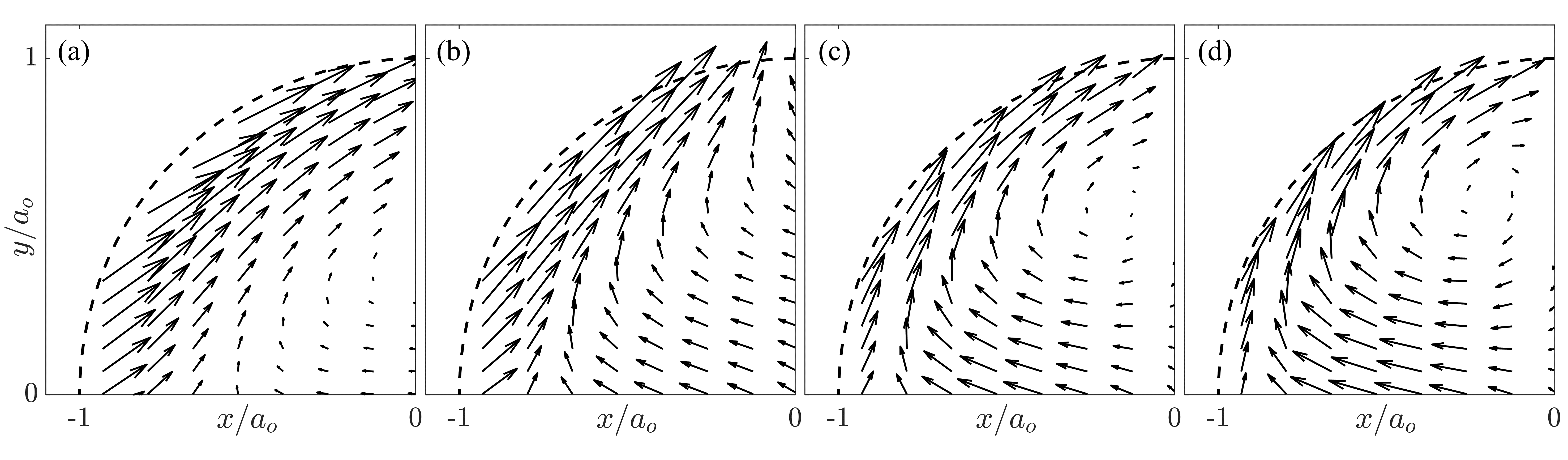}}
  \caption{Temporal evolution of the internal circulation of the drop for Case II. Velocity vectors inside the drop are shown at different times: (a) $t / \tau_{def} = 0$, (b) $t / \tau_{def} = 0.8$, (c) $t / \tau_{def} = 2.6$, and (d) $t / \tau_{def} = 5.3$. Fewer and larger vectors are displayed for visualization purposes.}
\label{Fig.TransientVector}
\end{figure*}

We now shift our focus to the dynamic response of the velocity at the interface of the drop. Figure \ref{Fig.TransientVector} shows the velocity vectors at different instants of time. We focus here on Case II, as this case still corresponds to a nearly spherical shape but with slightly larger velocity magnitudes. Note also that in Fig. \ref{Fig.TransientVector} less vectors are shown with larger scale, for visualization purposes. The time instant $t$ is made dimensionless based on the time scale of the deformation of the drop $\tau_{def}$. Clearly, the internal structures of the velocity field evolve with time. Electrohydrodynamic flows originate at the interface due to a discontinuity in electric stresses (see the stress balance given by Eq. \ref{Eq.StressBalance}). The velocity firstly appears as a tangential component at the interface ($t/ \tau_{def}=0$), followed by an increase in the radial component within the drop. The magnitude of the vectors in the bulk of the drop increases with time for later values of time  when $t / \tau_{def} =0.8$ and $t / \tau_{def} = 2.6$ until a clear vortical structure is identified on the right top corner of the drop at $t / \tau_{def} = 5.3$ (d) after which the permanent regime is established. Although we show the transient velocity vector only for Case II, a similar qualitative behavior has been observed for all other cases, regardless the magnitude of $E_o$.

We now compare our measurements to the closed-form solution of the dynamic response of the velocity field, presented by \citet{Esmaeeli2011}. We focus here only on the internal EHD circulation of the drop, for which the following streamfunction is proposed: 

\begin{equation}
    \psi = (Ar^3+Br^5) \sin ^2\theta \cos \theta
    \label{Eq.PSI}
\end{equation}

\noindent with $\mathbf{v_r} = \frac{1}{r} \frac{\partial \psi}{\partial \theta}$ and $\mathbf{v_{\theta}} = - \frac{\partial \psi}{\partial r}$. $A$ and $B$ carry the time-dependency of the velocity field:

\begin{multline}
 A = \frac{V_m}{a} + \frac{3}{2} [\frac{v_s}{(R+2)^2 a}][\frac{3M+2}{(19M+16)(2M+3)}] \\ f_T \exp{(-t/ \tau)}
 \label{Eq.A}
\end{multline}

\begin{multline}
 B = - \frac{V_m}{a^3} - \frac{9}{2} [\frac{v_s}{(R+2)^2 a^3}][\frac{1}{19M+16}] \\ f_T \exp{(-t/ \tau)}
 \label{Eq.B}
\end{multline}

\noindent where $\tau$ is a capillary time scale that governs the dynamics:

\begin{equation}
\tau = \frac{\mu a_o }{\gamma} \frac{(19 M+16)(2M+3)}{(40 M+40)}
\label{Eq.TauCapModified}
\end{equation}

\noindent that is modified to take the viscosity ratio $M$ into account. Note that $\mathbf{v_r}$ and $\mathbf{v_{\theta}}$ obtained from this theory eventually converge tho the LDM prediction at steady state when $t \rightarrow \infty$.

\begin{figure*}
  \centerline{\includegraphics[width = \textwidth]{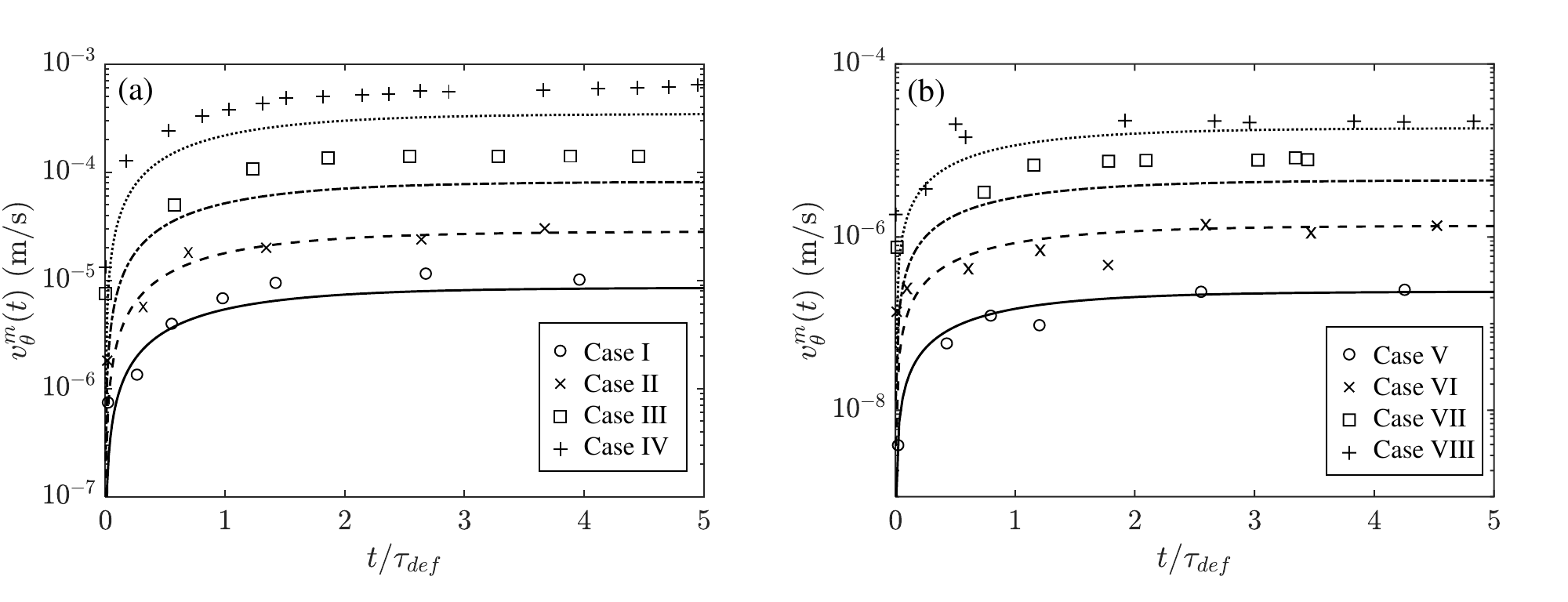}}
  \caption{Dynamic response of the maximum component of the tangential component of the velocity at the interface of the drop, $v_{\theta_i}^m$ for Cases I - IV (a) and Cases V - VIII (b). Markers indicate the experimental measurements, while the solid, dashed, dash-dotted, and dotted lines refer to $\mathbf{v_r} = \frac{1}{r} \frac{\partial \psi}{\partial \theta}$ and $\mathbf{v_{\theta}} = - \frac{\partial \psi}{\partial r}$ for each case, where $\psi$ is given by Eq. \ref{Eq.PSI}.}
\label{Fig.TransientInterfacialVelocity}
\end{figure*}

The dynamic response of the $v_{\theta_i}^m$ is shown in Fig. \ref{Fig.TransientInterfacialVelocity} (markers) for all the cases investigated. Conveying the point made from the investigation of the transient velocity vectors shown in Fig. \ref{Fig.TransientVector}, that the EHD circulation begins firstly as a tangential component at the interface when the electric field is applied, it is foreseeable to have an exponential growth of $v_{\theta_i}^m$ over time until the steady state condition is reached. By a direct comparison of our measurements to the transient analytical solution of $\mathbf{v_r}$ and $\mathbf{v_{\theta}}$, shown as the lines in Fig. \ref{Fig.TransientInterfacialVelocity}, it is observable that the streamfunction given by Eq. \ref{Eq.PSI} is capable of properly describing the dynamic response of the tangential velocity at the interface of the drop, at least when regarding its maximal value $v_{\theta_i}^m$. However, we use here the usual definition of the capillary time scale as the governing parameter of the dynamic response so that $\tau = \tau_{def} = \mu a_o / \gamma$, which seemed to be a better fit to our measurements. Apparently, the viscosity ratio $M$, although important in the dynamic response of the shape of the drop, has minor effect on the velocity at the interface.  As the transient solution proposed by \citet{Esmaeeli2011} is based on the same assumptions than the LDM theory regarding the steady-state condition, it is expected to see an under prediction of our measurements for the cases that correspond to larger deformations. We confirm here the critical value of $Ca_E \approx 0.1$, below which the analytical solutions for both the transitory and permanenet regimes are applicable. It is also noteworthy that a monotonic growth is observed for all cases investigated. This is in corroboration with other remarks from Esmaelli \& Behjatian (2020) \cite{Esmaeeli2020} who argue that an oscillatory dynamic response can be expected for less viscous fluids, which is not the case of the current measurements.

\section{Conclusions}
\label{Section.Conclusions}

Experimental measurements of the electrohydrodynamic (EHD) flows inside a neutrally buoyant drop have been reported for the first time. Special focus is given for the tangential component of the velocity at the interface $\mathbf{v_{\theta_i}}$, for which the dynamic response is also investigated. By using two leaky dielectric oils, namely, Silicone and Castor oils, both the oblate and prolate configurations were investigated. In addition, the strength of the constant electric field $E_o$ spanned from 0.125 to 1.75 kV/cm, thus covering both the small-deformation limit ($Ca_E \ll 1$) and also larger deformations. The internal circulation at the plane of symmetry of the drop was measured in an Eulerian referential using particle image velocimetry (PIV) and high-speed shadowgraph.

While a similar overall behavior of the tangential velocity $\mathbf{v_\theta}$ and $\mathbf{v_r}$ in the permanent regime was observed compared to the analytical prediction, the LDM theory under predicts our measurements when the drop adops an ellipsoidal shape. However, even within the small-deformation limit, some deviations in $\mathbf{v_\theta}$ are observed due to the deformation of the drop, which is not accounted for by the LDM model. We confirmed that the electrohydrodynamic flows are indeed driven by the electric stress jump at the interface, as described by the balance equation $\mathbf{n} \cdot [(\mathbf{T} - \mathbf{\hat{T}})+(\mathbf{T_{el}} - \mathbf{\hat{T}_{el}})] = \gamma \mathbf{n} (\nabla \cdot \mathbf{n})$. The velocity starts as a tangential component at the interface, followed by an increase in the radial component within the drop, leading to internal circulation. The dynamic response of the velocity field is then described by the capillary time scale $\tau_{def} = \mu a / \gamma$, for which current theories available in literature have a fair agreement. We propose a critical electric capillary number of $Ca_E \approx 0.1$, below which the LDM approach adequately represents the both the transitory and quasi-steady dynamics of the velocity field. However, we stress that no analytical solution is present for the case of larger deformations, thus the novel methodology presented here offers significant improvements towards the capability of understand the internal EHD flow of a neutrally buoyant drop, as it offers no restrictions to the experimental conditions employed, thus providing proper measurements to cases where no analytical solutions are available. 

\section{Acknowledgements}

This research was funded by the Normandie Region, under the grant 22E05147. We also acknowledge the technical support of Franck Lefebvre, Gilles Godard, and Said Idlahcen from CORIA during the implementation of the experimental set-up. The measurements of the dielectric properties were conducted by Nicolas Delpouve and Laurent Delbreilh from the GPM (Groupe de Physique des Mat\'eriaux) of the UFR Sciences et Techniques de Rouen, whose contribution is also hereby acknowledged.

\appendix

\section{Correction of image distortion}\label{appA}

\begin{figure*}
  \centerline{\includegraphics[width = \textwidth]{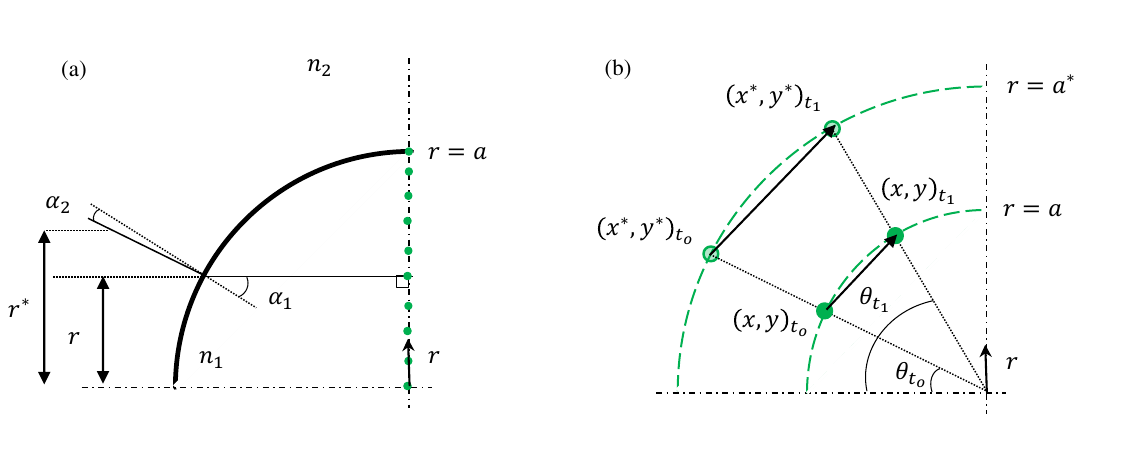}}
  \caption{Scheme of the correction of interfacial velocity based on image refraction, considering Case I. (a) Lateral view of the drop showing the light refraction at one instant of time. (b) Frontal view of the drop showing sample particles at the interface at two instants of time $t_o$ and $t_1$. The $*$ symbol denotes the quantities measured directly from the distorted image.}
\label{Fig.Refraction}
\end{figure*}

The RhodamineB particles emit light (wavelength of 575 nm) when illuminated by the laser sheet. While this happens at the axis of symmetry of the drop (refractive index $n_1$), the light will be refracted once it crosses the interface of the drop with the medium (refractive index $n_2$), as shown in Fig. \ref{Fig.Refraction} (a), according to the Snell's law of refraction:

\begin{equation}
    n_1 \sin \alpha_1 = n_2 \sin \alpha_2
    \label{Eq.Snell}
\end{equation}
    
\noindent where $\alpha$ is the angle formed between the light beam emitted from the particles and the vector normal at the interface of the drop. Since $n_1 \neq n_2$, the image captured by the CCD camera will be distorted. Considering Case I, where $n_2>n_1$, it follows that $\alpha_1>\alpha_2$, thus the particles will appear to be closer to the interface as they actually are with $r^* > r$, where the $*$ symbol denoted the virtual position, obtained from the distorted image.

Let us consider the velocity at the interface:

\begin{equation}
    \mathbf{v_{\theta_i}^*} = \dfrac{\sqrt{\left( x_{t_1}^* - x_{t_o}^* \right) ^2 + \left( y_{t_1}^* - y_{t_o}^* \right) ^2}}{t_1 - t_o}
    \label{Eq.UncorrectedInterfacialVelocity}
\end{equation}

\noindent where $\mathbf{v_\theta^*}$ is the tangential component of the velocity at the interface, obtained directly from the PIV calculation, according to the scheme shown in Fig. \ref{Fig.Refraction} (b). The positions in Cartesian coordinates are: $x_{t_i}^* = a^* \cos \theta_{t_i}$ and $y_{t_i}^* = a^* \sin \theta_{t_i}$. The corrected interfacial velocity is therefore:

\begin{equation}
    \mathbf{v_{\theta_i}} = \dfrac{\sqrt{\left( x_{t_1} - x_{t_o} \right) ^2 + \left( y_{t_1} - y_{t_o} \right) ^2}}{t_1 - t_o}
    \label{Eq.CorrectedInterfacialVelocity}
\end{equation}

\noindent based on the corrected $x_{t_i}$ and $y_{t_i}$ positions. Note that at the interface there is no radial component of the velocity, $i.e.$ $v_r(r=a,r=a^*)=0$. One can consider $a,a^* \amalg \theta_{t_i} \vert r=a,a^*$ which leads to $\theta_{t_i}  \amalg n^* \vert r=a,a^*$, where $n^* = n_1 / n_2$.

By defining a correction factor for the interfacial velocity $f_i = \frac{v_{\theta_i}}{v_{\theta_i}^*}$ and considering small particle displacements - around twice the particle size in this example - one can approximate the particle displacement $\sqrt{\left( x_{t_1}^* - x_{t_o}^* \right) ^2 + \left( y_{t_1}^* - y_{t_o}^* \right) ^2}$ as the arc length obtained from the path of each particle in curvilinear coordinates:

\begin{equation}
    f_i \approx \dfrac{\left( \theta_{t_1} - \theta_{t_o} \right) a }{\left( \theta_{t_1} - \theta_{t_o} \right) a } \therefore f_i \approx \dfrac{a}{a^*}
    \label{Eq.Factor1}
\end{equation}

\noindent which implies that $f_i$ is simply the ratio between the virtual and real positions of the interface and can be obtained directly from Eq. \ref{Eq.Factor2}: 
\begin{equation}
    f_i \approx \lim_{\alpha_1 \to \pi/2} \tan(\alpha_1-\alpha_2)
    \label{Eq.Factor2}
\end{equation}

\noindent where 

\begin{equation}
    \alpha_2 = \arcsin{ \left( \dfrac{n_1}{n_2} \lim_{\alpha_1 \to \pi/2} \sin \alpha_1 \right)}
    \label{Eq.Alpha}
\end{equation}

\noindent we stress that such procedure considers fluids with similar refractive indexes ($n^* \rightarrow 1$) and may not be adequate for other systems of fluids.

\bibliography{refs}

\end{document}